# The two faces of the coin of Special Relativity

**Ibrahim M. Al Abdulmohsin**
Saudi Aramco, P.O.Box 13423, Dhahran 31311, Saudi Arabia

Email: abdulmim@aramco.com

**Abstract.** In the educational literature, it is rarely emphasized in modern physics textbooks that our definitions of space and time have to reflect their complete interdependence. Our intuitive methods of always picturing one-dimensional space as a sum of unit-length rods and of picturing one-dimensional time as consecutive ticks of a clock suggest subconsciously that changes in space never alter time and vice versa. In this paper, we present a compiled list of popular arguments against Special Relativity, and we show how our erroneous intuitive conviction that the definition of space is always independent of the definition of time and vice versa is indeed the driving force behind these arguments. Once interdependence between space and time is established within their respective definitions, it is shown how such arguments are quickly resolved. Interdependence between space and time is described in two different ways, which are mathematically different but physically equivalent. We show how by assuming a time-independent measurement tool for space, time should be defined in terms of space to reflect their interdependence and this yields Lorentz Transformation. If we assume a space-independent measurement tool for time, space should be defined in terms of time, and this yields a different set of equations. Next, it is shown how the two different faces of the coin of Special Relativity, while mathematically different, are essentially equivalent from a physical point of view, and that both yield the same laws of addition of velocity, length contraction and time dilation effects. The *apparent* differences in physical interpretations between the two sets of equations are attributed to the differences in the definitions of space and time used in both cases. In short, this paper makes four contributions to educational physics. It presents a simple proof to Lorentz Transformation with emphasis on physical interpretation throughout the proof. It shows how interdependence between space and time must sometimes be reflected in our definition of such terms, and how Lorentz equations can be explained by defining time in terms of space. Third, it presents simple answers to popular questions that are frequently raised against Special Relativity and/or Lorentz Transformation. Finally, it shows that Lorentz Transformation is not the only set of equations that can accurately describe Special Relativity, where at least one other set of equations can be constructed by defining space in terms of time instead.

Alternative mathematical model for special relativity

1. **Introduction**

In 1905, Einstein proposed his Special Theory of Relativity to resolve the apparent discrepancies between classical Galilean laws of motion and Maxwell's electrodynamics. His theory is built on the fundamental postulate that all laws of physics are exactly the same across all inertia frames. Thus, Maxwell's laws of electromagnetism, which imply a constant speed of light in vacuum, are applicable to all inertia frames. The implications of this theory were revolutionary as they drastically altered our understanding of space and time [1]. Earlier in 1904, Lorentz proposed a transformation of coordinate systems between inertia frames that replaced Galilean transformation and led to the same physical implication that Einstein proposed independently later. According to Lorentz Transformation, space and time coordinates of a frame of reference for a moving object relative to an observer can be translated into space and time coordinates of the observer's frame of reference using equations 1 and 2 below. These equations can be easily used to derive the law of addition of velocities, which is shown in equation 3. The length contraction and time dilation formulas are given in equations 4 and 5 respectively [2].

$$x = (1 - v^2/c^2)^{-1/2}(x' + vt'). \tag{1}$$

$$t = (1 - v^2/c^2)^{-1/2}(vx'/c^2 + t'). \tag{2}$$

$$x/t = (x'/t' + v)/(1 + x'/t' \cdot v/c^2), \tag{3}$$

$$\Delta x = (1 - v^2/c^2)^{1/2} \Delta x' \tag{4}$$

$$\Delta t = (1 - v^2/c^2)^{-1/2} \Delta t' \tag{5}$$

The case for STR has been demonstrated both theoretically and experimentally over the past century. Experiments that have verified the constancy of the speed of light include, but not limited to, Michelson (1887), Illingworth (1927), Joos (1930), Cedarholm (1958), Antonini (2005), Herrmann (2005) and others [3]-[8]. There have also been other experiments that supported the time dilation effect such as Rossi (1940) [9]. Despite its enormous success, however, criticism still arises against STR by many students and, sometimes, even instructors. Some arguments against STR, such as the Twin Paradox, are chiefly due to lack of understanding of STR and are outside the scope of this paper. Others, however, are, in fact, due to our erroneous intuitive interpretation of space and time to be completely independent of each other. It is the latter type of arguments that will be tackled in this paper. In this paper, we will show how such arguments can be resolved by explicitly defining space in terms of time or by defining time in terms of space. The two methods yield two different sets of equations for the transformation of coordinate systems between inertia frames of reference but they are essentially equivalent from a physical point of view as will be shown later. They are, in other words, nothing but two different faces of the same coin; the coin of Special Relativity.

In this paper, we will first describe some common arguments against the Special Theory of Relativity and/or Lorentz Transformation. We will present several conditions for a successful theory of space and time distortions based on well-established theoretical and experimental facts. After that, we will show how by defining time in terms of space, and using the conditions concluded earlier, we arrive at Lorentz Transformation. By defining space in terms of time, on the other hand, a different set of equations for the transformation between inertia frames is constructed, where both transformations yield, consistently, the same physical implications. Next, we will show how the arguments against STR and/or Lorentz Transformation are quickly resolved after interdependence between space and time is established. Finally, we end with important conclusions that have to be addressed in the educational physics literature.

## 2. Common arguments against the Special Theory of Relativity and/or Lorentz Transformation

*2.1 First Argument: Considering the original assumptions behind Lorentz Transformation (i.e. the standard configuration), why is time still dependent on the position of the moving object?*

In the $2^{nd}$ equation of Lorentz Transformation, the time dilation effect appears to be dependent on the position of the moving object $x'$. While STR implies that time flows at exactly the same rate across all points in the same inertia frame, the $2^{nd}$ equation appears to be stating quite the opposite. To reconcile it with STR, common derivations of the time dilation effect simply ignore the space parameter $x'$ in the $2^{nd}$ equation of Lorentz Transformation and assume it to be zero, which is unacceptable from a mathematical point of view. Thus, the first argument against Lorentz Transformation is the *apparent* discrepancy between equation (2) and equation (5) above. It will be shown later that in Lorentz Transformation, space is assumed to be measureable using a time-independent measurement tool, while time is defined in terms of space to reflect their interdependence. It is this particular choice of definitions that is reflected in both the $1^{st}$ and the $2^{nd}$ equations of Lorentz Transformation. Equation (5), on the other hand, is the time distortion effect when time is assumed to be measured using a space-independent measurement tool.

Alternative mathematical model for special relativity

*2.2 Second Argument: Using the time dilation and length contraction formulas, why still can't we deduce the law of addition of velocities?*

If an object *A* is moving at velocity *v* away from an observer, and if a second object *B* is moving away from object *A* at velocity *dx'/dt'*, then the observer should calculate a velocity *dx/dt* for object *B* that is given in equation (6), which according to Lorentz Transformation should be equivalent to equation (7). This is because the observer will substitute *dx'* with $(1/\lambda)dx'$, and *dt'* with $\lambda dt'$, where $\lambda = (1 - v^2/c^2)^{-1/2}$. This, of course, is not the case in Lorentz Transformation.

$$dx/dt = v + (\eta \cdot dx')/(\kappa \cdot dt') \quad (6)$$

, where $\eta$ and $\kappa$ are the length distortion and time distortion factors respectively.

$$dx/dt = v + (1/\lambda^2)\, dx'/dt' \quad (7)$$

The two arguments might seem independent at first sight, but they are, in fact, equivalent. If we assume that equation (5) replaces equation (2) in Lorentz Transformation, and after dividing equation (1) by equation (5), we will indeed obtain the law of addition of velocities given in equation (7). Thus, the two arguments so far are essentially one argument; namely why the time dilation effect seems to depend on the position of the moving object *x'*.

*2.3 Third Argument: Considering a photon that travels at the negative spatial direction in the standard configuration, why is the speed of light still constant when length contraction and time dilation effects cannot preserver its constancy?*

It is clear that if a photon is moving away from the observer in the positive spatial direction in the standard configuration, length contraction and/or time dilation is needed to preserve the constancy of the speed of light. This is because classical Galilean transformation would predict an observed speed of light that is larger than *c*, since it would be the sum of the two velocities *c* and *v*. Consequently, length contraction and/or time dilation is needed to make the speed of light constant across all inertia frames. However, neither length contraction nor time dilation can explain why the speed of light would still be constant when a photon moved in the opposite direction, where Galilean transformation would predict a speed *less* than *c* for that particular photon. Since Lorentz Transformation does indeed preserve the constancy of the speed of light in all directions as clearly illustrated in its resultant law of addition of velocities, and since its 2[nd] equation does not directly imply a time dilation effect as mentioned earlier in the pervious two arguments, then a different interpretation of the 2[nd] equation is needed to describe why the speed of light is still constant even if the photon is moving at the opposite direction. The answer to this argument lies in our definition of time as will be described later.

*2.4 Fourth Argument: Considering a photon that travels at the positive spatial direction in the standard configuration, why aren't the standard length contraction and time dilation formulas able explain the constancy of the speed of light?*

Considering a photon that travels at the positive spatial direction in the standard configuration in inertia frame $I'$ that is moving at velocity $v$ from inertia frame $I$, and assuming the time dilation effect given in equation (5), the distance traveled by the photon as observed by $I$ is given in equation (8), where $\eta$ is the length distortion factor. After dividing this equation by equation (5), we arrive at equation (9). Knowing that the speed of light is constant across all inertia frames, then $x/t = x'/t' = c$, which implies that the length distortion factor is the one given in equation (10). The resultant length distortion effect is clearly not equivalent to the standard length contraction effect given in (4).

$$x = \eta\, x' + vt \tag{8}$$

$$x/t = (1 - v^2/c^2)^{1/2} \eta\, x'/t' + v \tag{9}$$

$$c = (1 - v^2/c^2)^{1/2} c\eta + v \Rightarrow \eta = (1 - v/c)(1 - v^2/c^2)^{-1/2} \tag{10}$$

Similarly, if we assume that the length distortion effect is indeed the one given in equation (4), we arrive at a different time dilation effect that is given in equation (11). It is important to note at this stage that the time distortion factor given in (11) is the reciprocal of the length distortion factor in (10), which will be shown later to be a direct result of their complete interdependence and switch of definitions. So, the length contraction and time dilation effects given in (4) and (5) are not sufficient to preserve the constancy of the speed of light. Why is this the case despite the fact that Lorentz Transformation does indeed preserve the constancy of the speed of light as illustrated in its resultant law of addition of velocities? As will be shown later, when interdependence between space and time is established, the time dilation and length contraction effects as specified in equations (4) and (5) cannot be *both* held true *simultaneously*, since this implies that space and time are completely independent of each other. When the space contraction effect in equation (4) is held true, for instance, we are implicitly assuming that space is measured using a time-independent measurement tool, and thus, time must be defined in terms of space, which gives us Lorentz 2$^{nd}$ equation. It is indeed thought-provoking to see how a rigorous mathematical approach such as Lorentz Transformation yields an insightful physical implication as the complete interdependence between space and time.

Alternative mathematical model for special relativity

$$\Delta t / \Delta t' = \left(1 - v^2/c^2\right)^{1/2} (1 - v/c)^{-1} = (1 + v/c)(1 - v^2/c^2)^{-1/2} \qquad (11)$$

### 3. Interdependence between space and time.

Before we delve into the conditions for a successful theory of space and time distortions, it is important to illustrate how interdependence between space and time should be included in our definition of at least one of those two terms. As discussed earlier, the main driving force behind STR is to explain the deviations between our observed velocities and what we would otherwise predict using the classical Galilean laws of motion. These deviations might be interpreted as either a space distortion effect, a time distortion effect, or both. To simplify analysis at first, let us suppose that an observer measures a sudden increase in the speed of light in vacuum. This change in the speed of light can be explained as a space contraction effect, whereby the observer's unit space contracted and, thus, the distance traveled by a photon in a unit time relative to a unit space increased. Similarly, the observer could explain the increase in the speed of light as a time dilation effect, whereby the observer's unit time increased, and thus, the distance traveled by a photon per a unit time increased subsequently. Because the observer cannot by any means claim that one explanation is more accurate than the other, both explanations are equivalent from a logical point of view. Thus, they should both be equivalent from a physical point of view as well. Since space contraction could be equivalent to time dilation in explaining differences between observed velocities and predictions, space and time shall no longer be treated independently in relativistic mechanics.

Interdependence between space and time can be established in our *definitions* of those terms. In general, both space and time can be defined with respect to real physical events or they could be defined using the readings of hypothetical measurement devices. For instance, an observer could send a beam of light from one specific point in space to another specific point and define the elapsed period as a unit time. This definition is a definition in terms of events, whereby a unit time is the period a photon *starts* moving from a particular point in space until it *arrives* at another particular point in space. Clearly, to define time in terms of events is to provide it with a space-dependent definition, where a change in space alters time immediately. Note, for instance, in the previous definition that if the distance between the two points changes, or equivalently the definition of a unit space changes, our definition of a unit time changes as well. If, on the other hand, time is *defined* using a *space-independent* measurement tool, which we shall denote as a *clock* at an observer's hand, time is independent of space. In the latter case, if the distance between the two points in space changes, the elapsed period changes but this latter change is

Alternative mathematical model for special relativityAlternative mathematical model for special relativity interpreted as a change in distance, not in unit time since a unit time is measured by a clock instead of being measured by events. Conversely, if unit space is defined as the distance a photon traverses from a particular temporal event, i.e. a specific clock reading, to another temporal event, i.e. a second clock reading, then changes at the rate of flow of time will change our unit space as well. Thus, by defining space in terms of events, a time-dependent definition of space is actually constructed. To define space independently of time, an observer has to define a unit space without referring to events. For instance, an observer could define a unit space as the length of a specific *rod*.

What are the physical implications of our definitions of unit space and unit time? From the previous analysis, it is clear that we have four theoretical combinations of how to define space and time. The first combination is to define space using a time-independent measurement tool, i.e. a rod, and to define time similarly using a space-independent measurement tool, i.e. a clock. The second combination is to define time in terms of spatial events while defining space independently of time. The third combination is to define space in terms of temporal events and to define time independently of space. The last option is define both space and time in terms of temporal and spatial events respectively. This last option is immediately discarded since it leads to a deadlock where neither space nor time can actually be defined. The first option is also discarded because it is not applicable to all physical phenomena. For instance, in order for an observer to measure space and time coordinates of a non-stationary object, the observer has the option of defining space using a rod, in which case elapsed time is *defined* as the time the object moves from one particular point in space to another point and, therefore, time is defined in terms of spatial events. The observer, also, has the second option of defining time using a clock, in which case the distance traveled by the object is *defined* to be the distance traversed in a specific period of time and, therefore, space is defined in terms of temporal events. Thus, only for stationary physical phenomena can an observer consistently define both space and time using a rod and a clock respectively. Establishing interdependence between space and time, on the other hand, is consistent and applicable to all physical phenomena including stationary ones. This conclusion that independence between space and time is only consistent for stationary physical phenomena while interdependence between space and time is necessary to measure the coordinates of non-stationary phenomena can, in fact, be shown to be a direct implication of Lorentz Transformation and/or the alternative transformation that will be proposed later.

To sum up, interdependence between space and time has to be established by either defining space in terms of time or by defining time in terms of space in order to consistently account for all physical phenomena. In addition, a deviation in observed velocities from classical

Alternative mathematical model for special relativity

Galilean laws of motion can be explained as either a time distortion effect, a space distortion effect, or a combination of both. Nonetheless, due to the relative motion between inertia frames, the options of explaining deviations between observations and the Galilean Transformation using space distortion or time distortion are no longer completely interchangeable as will be shown in the next section. There are several conditions that have to be satisfied in order for any theory of space and time distortions to be logically and physically consistent.

### 4. Conditions for a successful theory of space and time distortions.

Any successful theory of space and time distortions must meet the following three conditions. These conditions are proved using well-established theoretical and experimental facts. In the subsequent sections, Einstein's clock synchronization is assumed.

*4.1 First Condition: The transformation of coordinate systems of inertia frames must be linear and symmetric.*

The transformation of coordinate systems of inertia frames must be symmetric, and, it must be linear. A symmetric transformation is needed because of the principle of relativity of velocities, which states that if an inertia frame *I'* is moving at velocity *v* relative to inertia frame *I*, then it is equally accurate to say that inertia frame *I'* is stationary while inertia frame *I* is moving at velocity *–v* relative to inertia frame *I'*. A linear transformation is needed because every spatial or temporal segment (i.e. a distance in space or a period in time) can be logically broken down into smaller segments. The sum of the transformation of the smaller segments has to be equal to the transformation of the original segment in order for the transformation to be logically consistent. Thus, in mathematically precise terms, the transformation must be of the following form:

$$\begin{pmatrix} x \\ t \end{pmatrix} = \begin{pmatrix} A(x' + bt') \\ D(kx' + t') \end{pmatrix} = \begin{pmatrix} A & Ab \\ Dk & D \end{pmatrix} \begin{pmatrix} x' \\ t' \end{pmatrix}. \tag{14}$$

*4.2 Second Condition: The transformation of coordinate systems must predict a constant speed of light in vacuum across all inertia frames.*

The case for the constancy of the speed of light has been demonstrated theoretically in Maxwell's equations of electromagnetic radiations. According to Maxwell's equations, the velocity of electromagnetic radiations, such as light, does not depend on the velocity of the object emitting those radiations. Thus, the velocity of light emitted by a rapidly moving body is equal to the velocity of light emitted by a stationary body, relative to an observer. Because the laws of physics are applicable to all inertia frames, all inertia frames must detect the same phenomenon;

Alternative mathematical model for special relativity i.e. the constancy of the speed of light. Thus, the velocity of light must not depend on the velocity of the emitting object, nor on the velocity of the observer. The case for the constancy of the speed of light has also been established empirically in numerous experiments as mentioned earlier. In mathematical terms, equation (14) implies the following law of addition of velocities:

$$x/t = (A/D)(x'/t' + b)/(kx'/t' + 1). \qquad (15)$$

In equation (15), the constancy of the speed of light implies that if $x'/t' = c$, then $x/t = c$. Similarly, if $x'/t' = -c$, then $x/t = -c$. These conditions lead us to the following equation.

$$b = kc^2. \qquad (16)$$

Next, it can be shown that equation (16) implies

$$A = D. \qquad (17)$$

*4.3 Third Condition: The time distortion factor has to be equal to the reciprocal of the length distortion factor in the special case where the observed event is in the inertia frame of the moving observer.*

The third condition for a successful theory of time and length distortions is having the time distortion factor equal to the reciprocal of the length distortion factor *in the special case where the observed event is in the inertia frame of the moving observer*. The proof of this condition is straightforward. Assume an observer *B* is moving at velocity *v*, relative to an observer *A*. If observer *A* measures a distance traveled by *B* to be equal to $x_0$, then *B* must have measured the distance to be equal to $n\ x_0$, where *n* is the length distortion factor. Because both *A* and *B* are measuring the same velocity relative to each other, *A* must measure a time $t_0$ that took object *B* to travel distance $x_0$, while *B* must measure a time $n\ t_0$ for the same period of time. Therefore, a unit distance in a moving inertia frame of reference, whose velocity is *v* relative to an observer, is observed as *n* unit distance by the observer. On the other hand, an *n* unit of time in the moving inertia frame of reference is observed as a single unit of time by the observer. Thus, the length distortion factor and the time distortion factor must be reciprocal to each other in the special case mentioned above. In mathematical terms, using equation (14), it can be shown that:

$$x = A(1 - bk)x' + bt. \qquad (18)$$

We also know that equation (19) must hold true, where $\eta$ is the length distortion factor.

$$x = \eta x' + vt \qquad (19)$$

Equating equations (18) with (19), it is shown that the length distortion factor is given by equation (20).

$$\eta = A(1 - bk). \qquad (20)$$

a

And,

$$b = v. \tag{21}$$

From equations (16), (17) and (21), the transformation in (14) must be of the form given in (22), and the resultant law of addition of velocities must be the famous law given earlier in equation (3). For simplicity and because all velocities are assumed to be constant in the standard configuration, $x/t$ will be denoted as $w$ while $x'/t'$ will be denoted as $u$, as shown in equation (23).

$$\begin{pmatrix} x \\ t \end{pmatrix} = A \begin{pmatrix} x' + vt' \\ v/c^2 \cdot x' + t' \end{pmatrix} \tag{22}$$

$$w = (u + v)/(1 + uv/c^2) \tag{23}$$

The law of addition of velocities in (23) can be rearranged as shown in equation (24). Equating equation (24) with equation (6), we arrive at equation (25) that relates the length and time distortion factors in the utmost general case where the observed event is *not necessarily* in the inertia frame of the moving observer $I'$. In equation (25), $\eta$ and $\kappa$ are the length distortion and time distortion factors respectively.

$$w = v + (1 - wv/c^2)u \tag{24}$$

$$\eta/\kappa = (1 - wv/c^2) \tag{25}$$

In the special case where the observed event is in the inertia frame of the moving frame $I'$, $w = v$, and, thus, using equation (25) and the third condition, we arrive at the famous length contraction and time dilation factors given in equation (4) and (5) respectively. This, however, is only a special case, and it does prove the most important results of STR. As a result, space *coordinates* contract and time *coordinates* dilate by the factors given in equations (4) and (5) when an inertia frame moves with a velocity $v$ relative to an observer because coordinates are, by definition, stationary relative to an observer. However, the transformation itself is not yet complete because light is never in the inertia frame of the moving frame $I'$, and it is the speed of light that we need to hold constant. The subsequent section will provide two different transformations of coordinate systems that account for the general case where an observed event is not necessarily in the moving inertia frame $I'$. Both transformations predict a constant speed of light in vacuum across all inertia frames by either defining space in terms of time or, alternatively, by defining time in terms of space, where the latter transformation is the famous Lorentz Transformation. The two transformations complete equation (22) by specifying the value of the coefficient $A$ in both scenarios.

Alternative mathematical model for special relativity

### 5. Completing the transformation: When space is defined in terms of time.

It is clear by now that when an inertia frame *I'* is moving at velocity *v* relative to inertia frame *I*, space and time *coordinates* of *I'* are distorted in *I* by the famous length contraction and time dilation formulas given in equations (4) and (5). However, the result is incomplete yet since it does not account for the general case where observed events are not necessarily stationary with respect to *I'* inertia frame. This has been mathematically illustrated in our discussion of both the third and fourth arguments as well as our discussion of equation (25). Since our objective is to arrive at a consistent transformation that preserves the constancy of the speed of light, the general case where observed events are not necessarily stationary with respect to *I'* inertia frame has to be accounted for. In this general case, we assume that all velocities, including *v* and *u*, are held constant. Thus, we know that equations (26) and (27) must hold true, where *u* and *w* are the velocities specified in equation (23). It is vital to keep in mind that *x*, *t*, *x'*, and *t'*, are *not necessarily* equivalent to $\Delta x$, $\Delta t$, $\Delta x'$, and $\Delta t'$ that are specified in the length contraction and time dilation formulas in equations (4) and (5). For instance, space coordinate *x* is always different from $\Delta x$, while *x'* and $\Delta x'$ can be equivalent. To illustrate this, consider the case where a photon is emitted away from the moving observer *I'*. In this particular scenario, what we actually denote as $\Delta x$ does not mean the position of light with respect to *I*, but the distance between the observer in *I'* (which is normally represented as the origin of *I'*) and the photon as observed in inertia frame *I*. The space coordinate *x* is clearly different from $\Delta x$, while *x'* is equivalent to $\Delta x'$.

$$x = wt \tag{26}$$

$$x' = ut' \tag{27}$$

To complete the transformation, we will first assume that time is independent of space. Thus, time is assumed to be a clock that is held at each observer's hand in both *I'* and *I*, and time distortion is defined to be the difference in measurement between the two clocks. Because the clocks are not affected by what is being observed, i.e. time is not defined in terms of events, and because the clocks are stationary relative to each observer (i.e. they are in inertia frames *I* and *I'* respectively) the time distortion factor is fixed and it should be equal to the time dilation value given in equation (28) as shown earlier in our discussion of equation (25).

$$\kappa = \left(1 - v^2 / c^2\right)^{-1/2} \tag{28}$$

Because time is assumed to be independent of space, space has to be defined in terms of time to reflect their interdependence as discussed earlier. This is illustrated in equations (26) and (27). By dividing equation (26) by equation (27), we arrive at equation (29). The value given in equation (29) is again *not* the space distortion factor since it is not equal to $\Delta x / \Delta x'$.

Alternative mathematical model for special relativity

$$x/x' = (w/u) \cdot t/t' = (1 - v^2/c^2)^{-1/2}(w/u) \tag{29}$$

From equations (19) and (29), we know that equation (30) holds true, where $\eta$ is the length distortion factor.

$$(1 - v^2/c^2)^{-1/2}(w/u) x' = \eta x' + vt \tag{30}$$

By dividing equation (30) with $t'$, and using equation (27), we know that equation (31) holds true. Thus the length distortion factor $\eta$ should be the one given in equation (32).

$$(1 - v^2/c^2)^{-1/2} w = \eta u + v(1 - v^2/c^2)^{-1/2} \tag{31}$$

$$\eta = (1 - v^2/c^2)^{-1/2}(w - v)/u = (1 - v^2/c^2)^{-1/2}(1 - wv/c^2) \tag{32}$$

As a result, in the case where time is assumed to be measured using a clock that is independent of space and where space is defined in terms of time, the transformation in (22) can be completed as shown in equation (33). This transformation can be rearranged into the form proved earlier in equation (22) by setting the coefficient $A$ to be equal to $(1 - vw/c^2)(1 - v^2/c^2)^{-3/2}$.

$$\begin{pmatrix} x \\ t \end{pmatrix} = (1 - v^2/c^2)^{-1/2} \begin{pmatrix} (1 - vw/c^2)x' + vt' \\ t' \end{pmatrix}$$
$$= (1 - vw/c^2)(1 - v^2/c^2)^{-3/2} \begin{pmatrix} x' + vt' \\ v/c^2 \cdot x' + t' \end{pmatrix} \tag{33}$$

The transformation in (33) does not imply that the *coordinates* of a given event ($x'$, $t'$) relative to a rest frame ($x$, $t$) depend on the velocity of the third inertia frame ($x''$, $t''$), since they clearly have no causal relationship. However, equation (33) states that differences in length and time measurements *of a particular event ($x''$, $t''$)* depend on two velocity vectors: the velocity of the event with respect to the rest frame ($x$, $t$) and the velocity of the event with respect to the moving frame ($x'$, $t'$). In other words, space is defined in terms of events as illustrated earlier in a previous section. This provides the necessary generalization.

The physical interpretation of the transformation in (33) is very simple and straightforward. To illustrate it, let us consider the scenario of having two observers moving away from each other at a constant velocity $v$, and let us suppose that we threw several rods, where each rod is moving at a constant but unique velocity. The two observers record the measured length of each rod and announce their measurements to each other. The new transformation simply states that the ratio of each two length measurements by the two observers of the same rod is not constant even if the two observers are moving at a constant velocity with respect to each other. In fact, the ratio of each pair of length measurements of the same rod depends on the velocity of that particular rod relative to both inertia frame *I* and *I'*, and it is given in equation (34).

$$\Delta x / \Delta x' = (1 - v^2/c^2)^{-1/2}(1 - vw/c^2) = (1 - w^2/c^2)^{1/2}(1 - u^2/c^2)^{-1/2} \quad (34)$$

This is not a new result, however. Using the standard length contraction formula given in (4) alone, the previous scenario would be described using the if-then approach. If the rod is stationary with respect to (*x'*, *t'*), then (*x*, *t*)'s length measurement would be shorter than (*x'*, *t'*)'s measurement, and the space distortion effect is the famous length contraction formula given in equation (4). On the other hand, if the rod is stationary with respect to (*x*, *t*), then (*x*, *t*)'s length measurement would be *longer* than (*x'*, *t'*) by the *reciprocal* of the length contraction factor given in (4). Furthermore, if the rod is not stationary with respect to either (*x*, *t*) and (*x'*, *t'*), then the ratio of two measurement could be calculated by first calculating how each observer would announce his/her own length measurement. Nonetheless, such an if-then approach is clearly not mathematically concise. The transformation given in (33), on the other hand, incorporates all these scenarios into a single mathematically-concise equation. In the first case where the rod is stationary relative to (*x'*, *t'*), *w=v*, and *u=0*. Thus, we arrive at the well-known length contraction factor. In the second case where the rod is stationary relative to (*x*, *t*), *w=0*, and *u=-v*, which would imply that (*x*, *t*)'s measurement is larger by the reciprocal of the standard length contraction factor as expected. In addition, equation (34) specifies the ratio of length measurements in the third case where the rod is not stationary relative to (*x*, *t*) and (*x'*, *t'*), which is incorporated in the transformation as well. A more thorough discussion on the physical interpretation of this transformation will be provided in a subsequent section.

Previously, we have shown how a change in the speed of light can be logically described as either a distortion in space or a distortion in time, where the space distortion factor, if space distortion is chosen to describe the change in the speed of light, would be the reciprocal of the time distortion factor, if time distortion is chosen instead to describe the change in the speed of light. As discussed earlier, observers cannot by any means distinguish between the two different explanations since they are both logically correct in describing the change in the speed of light. Thus, they should both be equivalent from a physical point of view.

The exact same reasoning is applicable to how observers interpret discrepancies between their measurements of velocities on the one hand, and what they would predict using the Galilean law of addition of velocities on the other hand. In the case of the transformation of coordinate systems between inertia frames, we have shown that there are distortion effects to space and time measurements as follows. There is *standard* length contraction and time dilation effects by the factors given in equations (4) and (5) to the *coordinates* of the moving inertia frame *I'* when observed by inertia frame *I*. There is also *generalized* length distortion and time distortion effects to the apparent *positions and periods of moving events*, where the first effect is merely a subset of the second effect. This second effect was chosen to be described so far as a *standard* time dilation effect, since time is assumed to be the measurement of each observer's clock at hand, as well as a *generalized* space distortion effect as illustrated in equation (33). However, it can also be equivalently described as a *standard* length contraction effect, when space is defined to be the length of each observer's rod at hand, as well as a *generalized* time distortion effect. In the second explanation, the generalized time distortion factor is expected to be the reciprocal of the generalized space distortion factor shown in equation (34) just as the standard length contraction factor is the reciprocal of the standard time dilation factor. This alternative approach is what we commonly refer to as Lorentz Transformation.

### 5. Completing the transformation: When time is defined in terms of space.

When space is assumed to be measured using a time-independent measurement tool, which we have denoted as a *rod* at hand, time should be defined in terms of space to reflect their interdependence using equations (35) and (36), which are basically the same as equations (26) and (27). Since space distortion is simply the ratio in length measurement of each observer's rod at hand, the ratio is always fixed and it is equal to the standard length contraction formula given in equation (37). From equations (8), (35), (36), and (37), equation (38) holds true.

$$t = x / w \tag{35}$$

$$t' = x' / u \tag{36}$$

$$\eta = (1 - v^2 / c^2)^{1/2} \tag{37}$$

$$\begin{aligned} t / t' &= u / w \cdot (x / x') = u / w \cdot (\eta\, x' + vt) / x' \\ &= u / w \cdot (\eta + vt / x') = u / w \cdot (\eta + v / u(t / t')) \\ &= \eta\, u / w + v / w(t / t') \end{aligned}$$

$$t / t' = \eta\, u /(w - v) = (1 - v^2 / c^2)^{1/2} u /(w - v)$$

$$t / t' = \kappa = (1 - v^2 / c^2)^{1/2} (1 - wv / c^2)^{-1} \tag{38}$$

It is important to note that the *generalized* time distortion effect shown in equation (38) is indeed the reciprocal of the *generalized* space distortion effect shown in equation (32), which is a result of their logical equivalence and switch of definitions as discussed earlier. Thus, the transformation of coordinate systems in this second equivalent explanation, when space is assumed to be measured using a time-independent measurement tool and time is defined in terms of space, is given in equation (39).

$$\begin{pmatrix} x \\ t \end{pmatrix} = \begin{pmatrix} (1-v^2/c^2)^{1/2} x' + vt \\ (1-v^2/c^2)^{1/2}(1-vw/c^2)^{-1} t' \end{pmatrix} \qquad (39)$$

It can be shown using the relativistic law of addition of velocities in equation (23) that equation (40) holds true for all values of $v, u$.

$$(1-v^2/c^2)^{-1/2}(v/c^2 \cdot u + 1) = (1-v^2/c^2)^{1/2}(1-vw/c^2)^{-1} \qquad (40)$$

Thus, the transformation in (39) is equivalent to the transformation in (41).

$$\begin{pmatrix} x \\ t \end{pmatrix} = \begin{pmatrix} (1-v^2/c^2)^{1/2} x' + vt \\ (1-v^2/c^2)^{-1/2}(v/c^2 \cdot u + 1) t' \end{pmatrix} \qquad (41)$$

Also, the transformation in (41) can be rewritten as shown in (42), which, in turn, can be rewritten as shown in equations (43) and (44).

$$\begin{pmatrix} x \\ t \end{pmatrix} = \begin{pmatrix} (1-v^2/c^2)^{1/2} x' + v((1-v^2/c^2)^{-1/2}(v/c^2 \cdot x' + t')) \\ (1-v^2/c^2)^{-1/2}(v/c^2 \cdot x' + t') \end{pmatrix} \qquad (42)$$

$$\begin{pmatrix} x \\ t \end{pmatrix} = \begin{pmatrix} (1-v^2/c^2)^{-1/2} x' + v(1-v^2/c^2)^{-1/2} t' \\ (1-v^2/c^2)^{-1/2}(v/c^2 \cdot x' + t') \end{pmatrix} \qquad (43)$$

$$\begin{pmatrix} x \\ t \end{pmatrix} = (1-v^2/c^2)^{-1/2} \begin{pmatrix} x' + vt' \\ (v/c^2 \cdot x' + t') \end{pmatrix} \qquad (44)$$

Equation (44) is obviously nothing but the well-known Lorentz Transformation. Thus, Lorentz Transformation is the transformation of coordinate systems between different inertia frames when space is assumed to be measured using a time-independent measurement tool and time is defined in terms of space.

### 5. Physical interpretation of the two transformations.

In order to make the two transformations easier to understand, they are rewritten as shown below. Note that in this form, it is straightforward to see what the space and time distortion factors are, which are basically the coefficients of $x'$ and $t'$ respectively. It is clear from

Alternative mathematical model for special relativity the two transformations that the standard time distortion factor is the reciprocal of the standard space distortion factor, while the generalized time distortion factor is the reciprocal of the generalized space distortion factor.

$$\begin{pmatrix} x \\ t \end{pmatrix} = \begin{pmatrix} (1 - vw/c^2)(1 - v^2/c^2)^{-1/2} x' + vt \\ (1 - v^2/c^2)^{-1/2} t' \end{pmatrix} \quad \text{(First transformation)}$$

$$\begin{pmatrix} x \\ t \end{pmatrix} = \begin{pmatrix} (1 - v^2/c^2)^{1/2} x' + vt \\ (1 - vw/c^2)^{-1}(1 - v^2/c^2)^{1/2} t' \end{pmatrix} \quad \text{(Lorentz Transformation)}$$

The physical interpretation of the two transformations is not as difficult as it may appear at first sight. They are basically the mathematical expressions when either space or time is defined in terms of events To illustrate it in an example, let us suppose that we have two observers at the origins of inertia frame $I$ and $I'$ in the standard configuration, where $I'$ is moving at a velocity $v$ relative to $I$. Let us also suppose that an object is also moving at a constant velocity relative to one observer. Following normal convention, we denote the velocity of the object relative to $I$ as $w$ and the velocity of the object relative to $I'$ as $u$. When both observers announce their velocity measurements, both observers will detect differences between their measurements of velocities and what they would predict using Galilean transformation. To explain the differences, the two observers have two options.

The first option is assume that time is measured independently of space. Thus, both observers will measure time using their own clocks at hand. As an example, let us assume that $v = u = 0.5\ c$. Consequently, $w = 0.8\ c$. Because space is defined in terms of time, each observer will measure the distance traveled by the object by multiplying the relative velocity by the elapsed time. Assuming $t = 1\ s$, it follows from the standard time dilation formula in equation (5) that $t' = \sqrt{3}/2\ s$. The distance the object has traveled during this period is calculated by each observer using relative velocity and elapsed time in each inertia frame (Note that the definition of space is dependent on time and is defined in terms of events):

$x = wt = 4/5\ m$

$x' = ut' = \sqrt{3}/4\ m$

Thus, the observer in inertia frame $I$ will measure the following distance between the origin of $I'$ and the object:

$\Delta x = x - vt = 4/5 - 1/2 = 3/10\ m$

Alternative mathematical model for special relativity

For the second observer, the distance between the origin of $I'$ and the object is $x'$ and it is equal to $\sqrt{3}/4 m$. Thus,

$$\Delta x / \Delta x' = 2\sqrt{3}/5$$

This value is not the standard length contraction formula because space is defined in terms of events. The standard length contraction formula would not yield a consistent answer. However, this value is exactly what we would predict using the generalized space distortion factor as shown below since time is defined independently of space.

$$\Delta x / \Delta x' = (1 - vx/c^2)(1 - v^2/c^2)^{-1/2} = 2\sqrt{3}/5$$

The alternative option is to assume that space is measured independently of time. In this case, each observer will measure time by dividing the traveled distance by the relative velocity of the object (Note that the definition of time is dependent on space and it is defined in terms of events). Let us assume that $x = 4/5$ m, then it follows from the standard length contraction formula in equation (4) that $x'$ is:

$$x' = (x - vt)(1 - v^2/c^2)^{-1/2} = \sqrt{3}/5$$

Each observer will calculate the elapsed time period by dividing the traveled distance by the relative velocity of the object, which gives us $t = 1$ s and $t' = 2\sqrt{3}/5$ s. Thus, the generalized time distortion factor is

$$t/t' = 5/(2\sqrt{3})$$

, which is the reciprocal of the generalized space distortion factor in the first option as expected, and can be calculated directly using equation (38). So, the two transformations are consistent with each other, and their apparent mathematical differences are attributed to the way space and time are defined in both options.

Aside from the mathematical calculations, what the two transformations say in physical terms is that what we denote by space and time measurements have to be in reference to a real physical event. In figure 1 below, we have three inertia frames $(x, t)$, $(x', t')$, and $(x'', t'')$. In this figure, an object $(x'', t'')$ is moving at a velocity $w$ in inertia frame $I$ and a velocity $u$ in inertia frame $I'$. By representing length contraction as the projection of a distance from one inertia frame to the other, we obtain a symmetric length contraction effect as stated in STR. In figure 1, therefore, an event that takes place in position $c$ in inertia frame $(x', t')$ also takes place in position $b$ in inertia frame $(x, t)$. However, at the exact time the first observer in $(x', t')$ sees the event in position $c$ in inertia frame $(x', t')$, the second observer in $(x, t)$ has not seen the event yet because the object is still in position $n$. After a specific period of time, the second observer will

Alternative mathematical model for special relativity also see the same event when the object reaches position *b* in inertia frame (*x*, *t*). The first transformation, where space is defined in terms of time, simply presents the difference in length measurements at an instant. In other words, at the exact time the first observer in (*x'*, *t'*) measures a distance |*ac*| of the object, the second observer measures a distance |*an*| of the same object, and the generalized space distortion effect is simply the ratio of these two measurements. Note that time is defined independently of space in the first transformation, while space is defined in terms of temporal events. The second transformation, i.e. Lorentz Transformation, presents the difference in time measurements when both observers see the same event, which takes place in positions *n* and *b* in inertia frames (*x'*, *t'*) and (*x*, *t*) respectively, which is clearly not equivalent to the standard time dilation formula. Note that time is defined in terms of space, while space is defined independently of time. So, the two transformation are essentially equivalent. The apparent differences in physical implications between the two transformations are merely attributed to the different definitions of space and time used in both transformations.

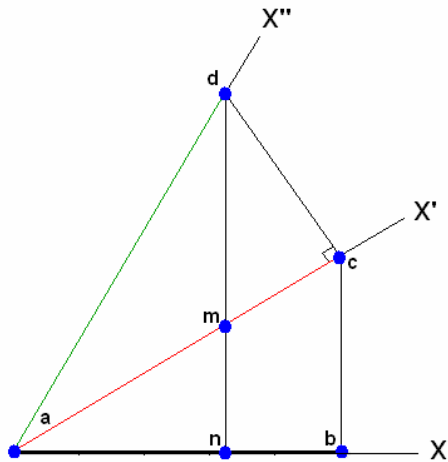

**Figure 1:** A graphical illustration of Special Relativity

### 6. Resolving the arguments.

*6.1 First Argument: Considering the original assumptions behind Lorentz Transformation (i.e. the standard configuration), why is time still dependent on the position of the moving object?*

The main fallacy in this argument is the assumption that both space and time should be defined independently of each other. As has been demonstrated throughout this paper, what we define by space and time measurements have to be in reference to real physical events. Thus, either space should be defined in terms of time or time should be defined in terms of space. The presence of the space coordinate in Lorentz $2^{nd}$ equation is merely an illustration of the fact that time in Lorentz Transformation is assumed to be defined in terms of space. If time is desired to be independent of space, another equivalent transformation can be constructed as shown in equation (33) by defining space in terms of time. In this alternative transformation, time dilation is indeed the one given in equation (5) while still preserving all conclusions of STR.

*6.2 Second Argument: Using the time dilation and length contraction formulas, why still can't we deduce the law of addition of velocities?*

The time and space distortion formulas predicted by Lorentz Transformation are, in fact, sufficient to deduce the law of addition of velocities. However, the time distortion factor predicted by Lorentz Transformation is not the *standard* well-known time dilation factor shown in equation (5). In fact, it is a *generalized* time distortion effect that is a result of defining time in terms of spatial events. As shown in equation (39), which was proven to be equivalent to Lorentz Transformation, the generalized time distortion effect is the one given in equation (38). Following the second argument in the beginning of this paper, it is easy to see that Lorentz Transformation is actually consistent with equation (6).

*6.3 Third Argument: Considering a photon that travels at the negative spatial direction in the standard configuration, why is the speed of light still constant when length contraction and time dilation effects cannot preserver its constancy?*

Because a photon is not in the inertia frames of both observers in $I$ and $I'$, the standard length contraction and time dilation formulas given in (4) and (5) respectively cannot be both held accurate at the same time. If space is assumed to be measured using a time-independent measurement tool, the time distortion factor is, in fact, the one given in (38), which may be a time dilation or a time contraction effect based on the direction of the velocity of the photon. If the photon is moving at the negative spatial direction, then we know that equations (45) and (46) both hold true, where $\eta$ is the space distortion effect, $x$ is the position of the photon in inertia frame $I$, and $x'$ is the position of the photon in inertia frame $I'$, and $k$ is the generalized time dilation effect.

$$x = \eta\, x' + vt \tag{45}$$

$$t = kt' \tag{46}$$

Because in Lorentz Transformation space is assumed to be measured using a time-independent measurement tool, $\eta$ is the standard length contraction effect shown in equation (4) as discussed earlier. After dividing (45) by (46) and knowing that the speed of light is constant across all inertia frames, we arrive at equation (47). Thus, the time distortion effect $\kappa$ is the one given in equation (47) for this particular scenario. This time distortion value is indeed the one predicted by the generalized time distortion effect in equation (38) where $w = -c$. Thus space and time distortion effects predicted by Lorentz Transformation do preserve the constancy of the speed of light even when a photon travels at the negative spatial direction.

$$-c = -(\eta/\kappa)\,c + v \tag{47}$$

$$\kappa = (1 + v/c)^{-1}(1 - v^2/c^2)^{1/2} \tag{48}$$

*6.4 Fourth Argument: Considering a photon that travels at the positive spatial direction in the standard configuration, why aren't the length contraction and time dilation effects able explain the constancy of the speed of light?*

The core of this argument are the differences between the space distortion factor in equation (10) and the space distortion factor in equation (4) when time is assumed to be measured using a space-independent measurement tool on the one hand, and between the time distortion effect given in equation (11) and the time distortion effect given in equation (5) when space is assumed to be measured using a time-independent measurement tool on the other hand. It has been shown throughout this paper that in the first scenario, the space distortion effect predicted by STR is not the standard length contraction factor given in equation (4) but the generalized space distortion factor given in equation (34). The latter equation does indeed predict the space distortion factor proved in equation (10) when a photon is traveling at the positive spatial direction. Similarly, the time distortion effect predicted by STR in the second scenario, which is given in equation (38), does also predict the time distortion effect proved in equation (11). Consequently, there is no mathematical or physical inconsistency in STR or Lorentz Transformation.

## 7. Conclusions.

A major source of misunderstanding of the Special Theory of Relativity or Lorentz Transformation is the erroneous intuitive conviction that space and time should be defined independently of each other. Once interdependence between space and time is established, many common arguments against STR or Lorentz Transformation are quickly resolved. Interdependence between space and time can be established using two different approaches, where both approaches define space and time with reference to real physical events. The first approach is to assume that time is measured using a space-independent measurement tool while space is defined in terms of time. The second approach is to assume that space is measured using a time-independent measurement tool while time is defined in terms of space, which is the approach used in Lorentz Transformation. Both approaches are essentially equivalent from a physical point of view despite their apparent different mathematical formulations. The apparent

Alternative mathematical model for special relativity differences between them are attributed to the different definitions of space and time used in both approaches.

Alternative mathematical model for special relativity